\documentclass[aps,prl]{revtex4}
\def\ee{\end{equation}}
\def\bea{\begin{eqnarray}}

\begin{document}

\title{Quanta and Qualia}

\author{Adrian \surname{Kent}}
\affiliation{Centre for Quantum Information and Foundations, DAMTP, Centre for
  Mathematical Sciences, University of Cambridge, Wilberforce Road,
  Cambridge, CB3 0WA, U.K.}
\affiliation{Perimeter Institute for Theoretical Physics, 31 Caroline Street North, Waterloo, ON N2L 2Y5, Canada.}
\email{A.P.A.Kent@damtp.cam.ac.uk} 

\date{August 2016, last revision May 2018} 

\begin{abstract}
I sketch a line of thought about consciousness and physics that
gives some motivation for the hypothesis that conscious observers
deviate -- perhaps only very subtly and slightly -- from 
quantum dynamics.   Although it is hard to know just how much credence
to give this line of thought, it does add motivation for a stronger and more
comprehensive programme of quantum experiments involving quantum observers. 

\end{abstract}
\maketitle
  
\section{Introduction}

I describe below my current stances on consciousness and its relationship
to physics, which have been strongly influenced in particular by James \cite{jamesautomata}.
Briefly, I am among those persuaded that there is a hard problem of
consciousness, and that in Chalmers' terms it appears to be strongly
emergent \cite{Chalmers2006-CHASAW} from, or in other words
inexplicable by, the currently understood laws of physics.  
This means I reject the counter-view that consciousness is a weakly emergent 
consequence of brain activity -- that is, a consequence that may initially seem
surprising but in principle is entirely explicable by aspects of neuroscience that are
reducible to known physical facts and laws.  

One strong reason for doing so is that a complete scientific model of 
consciousness appears to require, among other things, some sort of
data selection principle, characterising the subset of information 
about material substrates of consciousness (such as our brains) that
has correlates in conscious states (such as our conscious minds).
The known laws of physics are not of a type that allows any such
principle to be derived.   I enlarge on this below.  
Another way of framing the essential point is the conceivability argument for the logical possibility of a consciousness-free universe, 
materially identical to ours and following the same known laws of physics, but inhabited by ``philosophical zombies''
\cite{iepconscious, chalmers1996conscious}
who act as we do but are aware of nothing; I find this equally
persuasive. 
A further compelling argument is the knowledge argument
\cite{knowledge,iepknowledge}: someone can have knowledge of all the
known facts about and laws governing the material world and yet,
in the absence of first-hand experience, not know what particular
types of experience are like.    

The scale of the explanatory gap between the 
known laws of physics and conscious experience seems most evident to me when we take a 
cosmological perspective and consider consciousness in the
context of Darwinian evolution.   Our consciousnesses 
seem very well designed to maximize our survival chances;
they also seem designed to allow us to report and discuss
our conscious states with others.   None of this appears 
explicable by the known laws of physics \cite{jamesautomata,stanepi}.   

Of course, these positions are all highly debatable and 
have been criticised and defended by many. (
See e.g. Refs. \cite{chalmers1995facing,chalmers1996conscious,dennett1993consciousness,glynn2003anatomy,knowledge,iepknowledge,iephard,iepepi,stanepi,Chalmers2006-CHASAW} 
and references
therein for some expositions and reviews.)
My goal here is not to make new cases for them.
Nor do I want to suggest that they necessarily imply that there 
must be new physical laws explaining every aspect of consciousness. 
Consciousness poses formidable problems for physics, including explaining the brute fact of its
existence, qualitative and quantitative aspects of conscious experience, the 
relationship between consciousness and matter, the evolutionary or 
cosmological origins of consciousness, and the evolution of human
consciousness from some presumably primitive form.  
It is not clear that any of the standard positions on consciousness (e.g. \cite{sep-consciousness,iepconscious}) 
is necessarily the right starting point for resolving all these
problems, nor
does it seem realistic to expect any new proposal to tackle them all at once.
 
This leaves a risk of succumbing to a form of learned\footnote{in both senses of the 
term} helplessness, given a voluminous literature in which
every plausible argument is opposed by plausible counterarguments
and every interesting position has potentially insurmountable
difficulties.  A better alternative than intellectual paralysis
may be to accept our present framings of the problems
may be conceptually inadequate and to look for lines
of thought that might suggest different ways of thinking 
about the relationship between physics
(as presently understood) and consciousness. 

I suggest here that, in particular, current developments
in the foundations of physics could suggest possible ways of
acquiring new empirical evidence about the relationship between
physics and consciousness.  
First, I will sketch in a bit more detail the stances summarised above, 
and then try to say something about their potential implications for 
quantum theory and experimental tests involving conscious observers. 

\begin{enumerate}
\item Consciousness -- the collection of perceptions, sensations,
thoughts, emotions, thoughts about perceptions, and so on, that we
experience -- is a natural phenomenon.   
We say something about the world when we say that we are conscious, 
just as we do when we say that the Earth is roughly round and that
solid objects tend to fall towards it.   Saying that an individual's brain runs
algorithms that include models of the individual, or that their body
tends to respond in a relatively predictable way to stimuli, also
says something about the world -- but it does not say or logically
imply anything about their consciousness, including its existence.     

\item  One of the main goals of physics is to give compressed descriptions
of natural phenomena.\footnote{Some might say this
is {\em the} goal of physics.}   Physical laws reduce a very large set of data
to a much smaller set.   For example, Newton's laws of gravity and
of motion not only characterise and quantify how and when solid objects fall
towards the Earth, but give us a unified description that includes the
large-scale behaviour of liquids and gases, the motion of
celestial and terrestrial bodies, and laws governing tides and atmospheres.    
So it is a reasonable ambition for physics to look for a compressed,
lawlike description of consciousness. 

\item The only certain examples we have, our brains and nervous systems,
suggest that consciousness is intimately bound up with the
properties of matter.   So a reasonable ansatz, or starting point,
for a lawlike description of consciousness would be a relatively 
compressed set of rules from which we can infer that when a 
physical system is in state $S$ its consciousness is in 
state $C(S)$.   We certainly want to allow that $C(S)$ may be empty,
since we don't want to assume that every physical system is
conscious.\footnote{Nor do we want to exclude this possibility.}
We should also allow for the possibility of a physical
system having more than one separate consciousness, since a human family or 
a city appear to be examples, and perhaps even a single human or animal brain 
can be.   So really we should say ``its consciousness is in state
$C(S)$, or its consciousnesses are in states $C_i (S)$ for a list $i$
in some index set $I(S)$''.    For brevity we leave this implicit
below. 

\item By the admittedly high standards of successful laws of
physics,
we don't have any remotely satisfactory lawlike description of consciousness.
We can say the waking human brain is generally conscious, and that
specific types of consciousness -- visual imagery, or smell, or
formulating speech -- are associated with activities in various 
regions of the brain (generally many such regions for any given
activity).     These seem like raw observational data that 
any theory should aim to explain.   Perhaps, more charitably,
they could also be seen as steps towards high level laws in a high level
description, which should eventually be superseded by more fundamental
laws.   In physical terms they seem
roughly comparable to the observations
that stones fall to the ground, birds go up in the air when they
flap their wings hard and clouds tend to float around in the sky. 
That is, they are generally true, though slightly vague, statements about
quite complex physical systems.   The history of physics encourages us to try to 
describe the underlying phenomena better -- more completely, more
simply, more precisely -- by formulating simple and
precise mathematical laws governing a smaller range of more elementary  
objects or quantities.\footnote{This remains true even if one believes that 
some properties of consciousness are necessarily beyond the 
possible scope of physical explanation, so long as one accepts that 
there are at least {\it some} aspects that could be characterised by simpler
laws.   For example, one might consistently believe there is no hope
for a physical explanation of the nature of the sensory experience of
seeing red, and still hope for simpler laws characterising when
a physical system is conscious, or conscious of images, or even conscious 
of redness, and perhaps even for laws quantifying some information
theoretic measures of its consciousness.   The history of physics encourages us to try to formulate
mathematical descriptions for those aspects of natural phenomena 
that appear as though they may allow this, even if other aspects 
presently resist description.   For example, investigating the
biomechanics and neurology of animals was worthwhile even in 
an era where there seemed no obvious prospect of 
understanding the nature or origin of life.}

\item The sort of law suggested so far is consistent with
consciousness being an epiphenomenon \cite{iepepi,stanepi}, that is, 
having physical causes or correlates but no effects on the 
material world.     Suppose that the laws of 
physics are complete, or 
complete enough to describe physics in many regimes, including the 
behaviour of matter on Earth.   Or at least that they are completable, in the sense that 
there is an as yet undiscovered unified theory $T$ of the sort physicists
conventionally imagine.  That is, one that includes quantum
theory and gravity, and fully describes the dynamics of matter, fields and 
spacetime, perhaps also including a theory of initial conditions
and/or other constraints -- but which makes no reference to
consciousness.   And suppose, just to simplify the
language of the discussion, that $T$ allows a sort of effective
reductionism in many contexts, so that with appropriate modelling,
which in principle can be justified from the fundamental principles of
$T$, we can describe physical systems $S$ interacting with their
environment $E(S)$, modelled in a way derived from the laws of physics
encapsulated in $T$.   In shorthand, we say such systems $S$ follow
the laws of physics given by $T$.   
Now, some of these physical systems $S$ -- human brains, for example
-- have associated non-empty consciousnesses $C(S)$.  But by
(conventional) hypothesis, $S$ follows the laws of physics given by
$T$ whatever the form of its consciousness $C(S)$.   We 
don't need to know anything about $C(S)$ to predict the physical
behaviour of $S$, or any of its physical properties (other than
those of $C(S)$ itself).    Indeed, we don't even need to be aware
of the phenomenon of consciousness in order to predict the physical
behaviour of $S$.    

On the view given so far, then, a complete
understanding of the physics 
of $S$ involves understanding $T$, deriving the predictions that $T$
makes for $S$, and then adding, as an extra interesting detail, that 
$S$ has a particular (maybe empty) form of consciousness $C(S)$. 
This detail is generally time-dependent, our experiences tell us: 
as the physical state of $S$ changes over time, the consciousness 
$C(S)$ generally also changes.\footnote{Quite what sort of 
time-dependent story about $C(S)$ should emerge for  
conscious spatially extended objects $S$ is very unclear. 
It seems as though $C(S)$ should be associated
with the worldtube of $S$ rather than any fixed worldline, and 
then it seems that neither proper time or any other single time
parameter is adequate to characterise the dependence of $C(S)$. 
Given that our only empirical examples, our own consciousnesses,
seem (?) to be associated with a single time parameter, even
though our brains occupy appreciable spatial regions, 
we have no real basis for speculating further about more general
possibilities.   We thus tentatively file this under ``questions that 
might one day be addressable if there is real 
progress on a physical theory of consciousness''.}

\item However, if consciousness is an epiphenomenon, and its 
epiphenomenal association with the material world is described
by simple laws, then it is very hard to understand how and why we 
evolved to have rich consciousnesses that contain a great deal   
of data highly relevant to our survival.   Darwinian evolution
takes place in the material physical world.   If consciousness
hitches a free ride on that world, then there is no particular 
need {\it a priori} for evolutionarily successful creatures to
be conscious.   \cite{jamesautomata}  Even if they are, there is no need for their 
consciousnesses to contain data relevant to survival.   
We could equally well be agilely escaping a tiger while conscious of nothing,
or aware only of the fermion numbers of our patellae,
or any other physical variables associated with our material selves.
On the epiphenomenal view of consciousness, the laws of physics encoded in $T$ are all that is relevant
to our body and brain functions during the escape; they
are also all that is relevant to describing the evolution of those
body and brain functions over aeons that include successful and tragic
encounters with tigers by earlier generations.
All that we need to explain our evolved traits is that our
direct ancestors tended to be over-represented in the successful
encounters (et cetera).  

\item It is also very hard to understand how, if consciousness is
  purely an epiphenomenon, we can talk about the contents of our
conscious minds, listen to ourselves doing so, and feel that we 
accurately represented ourselves.     \cite{stanepiselfstult}

\item As if these problems for the 
epiphenomenal view of consciousness were not devastating enough, they
can be sharpened further.    \cite{jamesautomata}
Not only do our consciousnesses contain a great deal of information
about the world relevant to our survival, but the information is associated with 
qualitative types of experience that seem designed to encourage evolutionarily
advantageous behaviour.   It seems to be logically possible, and
arguably even natural, to think of epiphenomenal consciousnesses as
value neutral -- simply registering aspects of the exterior and 
interior world without associating any form of judgement on them. 
Indeed, some aspects of our own consciousnesses seem to be close
to neutral.   For example, we feel that many -- though by no means all -- 
visual images give us data without associating much aversion or
attraction.   However highly pleasant and
unpleasant sensations, pleasures and pains, play very significant
roles in our conscious lives.   Moreover, these roles {\em seem}
to be important to our survival.     By and large, the pleasures
seem associated with evolutionarily advantageous activities (food,
drink, friendship, bonding, raising of status, sex, $\ldots$), and the
pains with disadvantageous ones (raging thirst, injury, lowering of status, rejection, $\ldots$).   
Yet, on an epiphenomenal view, there seems no possibility of an 
evolutionary explanation for these correlations.   Evolution of 
our material selves explains that the   
laws of physics encoded in $T$ caused our brains and bodies to tend to  
seek out evolutionary advantageous activities and avoid
disadvantageous ones.   It does not then matter whether our epiphenomenal
consciousnesses find the former pleasurable and the latter painful,
or vice versa.    

\item So, consciousness is not an epiphenomenon.\footnote{Or else the laws
of physics and the laws of epiphenomenal consciousness were 
apparently designed together to give us the false and useless but pleasing
sense of being in conscious control of our actions.
I find it hard to take this possibility seriously. 
Epiphenomenalism {\em per se} does have advocates
\cite{iepepi,stanepi}. 
However, the relevant position here is stronger than standard
epiphenomenalism.
It is that there are succinct physical laws {\it and} 
separate succinct laws describing how the contents of consciousnesses
are connected epiphenomenally to their material substrates, 
and that the second types of law just happen to imply that
consciousnesses have the sense of controlling evolutionarily
relevant behaviour in evolutionarily favourable ways, even 
though in fact all the relevant behaviour is already physically
determined.   
Since almost all succinct epiphenomenal laws that one could imagine
do not have this property, and all of them are equally logically
consistent, this seems to me entirely implausible.   

It is worth adding here that even a very slight causal effect 
of consciousness on the material world -- if we could make sense
of such a hypothesis -- could allow the possibility of countering this specific objection.   
This is not because it would make it true that we are in conscious
control of our actions in the pleasing but naive way that we tend
to intuit: a very small effect would presumably not make much 
difference to this.   However, it might leave room for an
explanation of how the sense of conscious control and the 
associated causal physical mechanisms could have co-evolved,
over the history of life on Earth, through the slow accumulation 
of small effects.} 

\end{enumerate}

\section{Backward or onward?}

At this point, one really needs to pause and take a breath, because
the terrain is not going to become easier if one presses further.  
Following the logic of the argument so far, there should be 
a physical theory of consciousness, but it should not be 
an epiphenomenal theory.   But are there any other coherent options?
And even if there might possibly be,  how could they do any better in 
explaining the puzzles of the evolution of consciousness?   
Even if one is willing to dream up equations somehow trying to characterize
a dynamical interaction between conscious states and familiar material physical
states, would they not necessarily work equally well if we relabelled
painful states as pleasurable and vice versa?    

When an argument runs into such 
difficulties, one should question one's premises.   Perhaps 
the whole line of thinking about consciousness we have set out 
is just misguided?   Perhaps one of the other standard lines
of thought \footnote{See the references cited above for discussions.}
is more promising after all?  Well, perhaps.  A review
is beyond my scope here.  
But I'm not convinced: it seems to me they also end up
either falsely \footnote{I claim ``falsely'' is a statement of fact, not
  opinion.   Whether they are right or wrong, the sort of ideas
  sketched in this paper show that there {\it could} be a 
  testable scientific theory of consciousness.}  denying any possibility of scientific progress on the
hard problem or creating insoluble puzzles of their
own.   

At the very least, it is clear from the literature 
that each position on consciousness finds thoughtful
critics who believe they can identify deep problems.  
If every line of thought runs into deep problems, one 
should arguably pursue the one that offers most hope of
bringing new data.  I will now argue that the one I have set
out does at least suggest the possibility of experimental progress on
the problem of consciousness, and with that, the possibility 
of saying at least something more about how
consciousness evolved.   I don't see how to take it far 
enough to sketch any plausible conjecture
about a satisfactory solution to the problem of pain and pleasure.   
Still, even a small chance of experimental progress is worth
pursuing, especially given the huge implications.  
And if there is experimental progress, perhaps it will
bring conceptual and theoretical progress, in the new light
of which these puzzles might seem less daunting.  

People must, I imagine, once have thought it pointless to ask why stones always fall,
birds sometimes fly, and clouds generally float in the sky.
Those were just part of the definition of stones, birds, and clouds. 
It must have seemed useless to such people to speculate that we might
be able to understand all this falling, flying and floating better if
stones, birds, clouds and everything else in the natural world
turned out to be made up of smaller constituents. 
After all, even if they were, it must have seemed that we would just
be left asking essentially the same question: why do
stone-constituents fall (at least when assembled into stones)
whereas bird-constituents sometimes fly (at least when assembled into
birds), and so on.   In a sense, on this last point, they were right.
Even now, we {\it do} still ask why the laws of general relativity and quantum theory
hold and not others.   But even if the essence of the question is in
some sense still the same, its form has changed as our understanding
developed, from an obstinate gatekeeper seemingly preventing progress
to a faithfully helpful guide along the long path to modern
physics. 

So, let us continue.    

\section{Quanta and Qualia}

\subsection{Quantum theory and the brain}

The hard problem of consciousness was a problem when we believed the 
world was described by classical physics.\footnote{
This is meant as an intellectual point, not a 
historical statement about discussions of consciousness.    
The term ``hard problem'' was coined by Chalmers in 1995
\cite{chalmers1995facing,chalmers1996conscious}. 
Many earlier discussions of the mind-body problem cover some of the
same ground as Chalmers; the intellectual history is reviewed in 
Ref. \cite{chalmers1996conscious} and the other general references
cited earlier.   However, regardless of how precisely it was
recognised, by whom, or when, the problem {\em existed} in 
the era of classical physics.   To put it another way, if we
believed today that the world were described by classical physics,
we should still be discussing the hard problem.   
It is possible that quantum theory may ultimately
shed new light on consciousness, or vice versa.
It is even possible that at a fundamental level the 
formulation of quantum theory might turn out to require consciousness
as a physical primitive.   But it is not the case that quantum 
theory introduced the hard problem into physics.} 
It may still be a
problem if and when quantum theory and general relativity are superseded.
There is no compelling reason of principle to believe that 
quantum theory is the right theory in which to try to formulate a 
theory of consciousness, or that the problems of quantum theory must have
anything to do with the problem of consciousness. 

That said, physics is where it is.   Quantum theory is our 
best current fundamental theory.   It works extremely successfully
in describing microscopic physics and some aspects of macroscopic
physics.   But it still has problems.
One is the long-standing problem of finding some
description of objective reality consistent with quantum
theory.\footnote{Of course, this too is keenly debated. 
For some reviews and discussions, see for example
Refs. \cite{saunders2010many,schlosshauer2007decoherence,
fuchs2003notes, cushing2013bohmian}.}  
Another is that we cannot rigorously define physically relevant field
theories in four dimensions, even in Minkowski space.   
And, of course, we do not know how to unify quantum theory and general
relativity. 
 
In summary, despite all quantum theory's successes, there are still 
reasons to question whether it completely describes all of physics.
So, let us start by supposing that quantum theory applies pretty 
well to systems like human brains.   However, let us keep an open mind on whether 
it captures absolutely everything that physics can say about them -- since this
has certainly not been well tested -- and see where this takes us.  

\subsection{Qualia}

According to one popular line of thought (see e.g. 
\cite{lewis1929mind,schrodinger1992life,chalmers1996conscious})
our consciousnesses
can be thought of as composed of very large numbers of individual 
sensation-components, or qualia.   The analogy here is with 
(what was once) the atomic hypothesis: that matter, in
all its rich variety, can be understood as composed of various types 
of elementary objects, atoms, in various proportions and combinations.
Modern chemistry eventually led to the classification of the elements,
and hence the elementary atoms, and to the postulated understanding
of (macroscopic terrestrial) matter as combinations of atoms.  
Similarly, one might think, visual perceptions can maybe be 
understood as some combination of a finite number of colour
and relationship qualia, emotions as combinations of finitely
many elementary emotional qualia, and so on.   

There is absolutely no evidence in favour of
this qualia-as-atoms-of-consciousness 
model.   If consciousness is indeed something that
can be modelled in any scientifically familiar way, it could
be as a field, or a manifold.   It could also, of course, be 
that there is some mathematical model that looks nothing like
anything we have encountered in physics so far.  
Still, if we are going to speculate about the relationship of 
consciousness to the rest of physics at all, we need some language,
and the qualia model gives a useful way of thinking about how 
a connection might be made.    So we will use it, while emphasizing
that our tentative conclusions are meant to apply more generally. 
The same fundamental questions arise whether consciousness is
built from atomic qualia or described by some other 
quantities.  And, importantly, the same conclusions follow. 

\subsection{Qualia from quantum states: a cartoon}

Whatever our consciousnesses are, they are almost certainly not
identical to the physical states that give complete 
descriptions of our brains.   
Even on a classical description, the vast majority of the information 
encoded in a configuration space description of the 
brain's physical state does not appear to be carried by the 
associated consciousness.  
This is true even of the highly coarse grained descriptions
that arise in a higher level neuroscientific model of the brain.    
For example, we are not aware of whether or not neurons only involved in
governing unconscious processes are firing, nor are we aware
of most of the complex sequences of firings that produce conscious
images.   

The brain contains roughly $10^{11}$ neurons, firing on average
roughly $10^2$ times per second, giving roughly $10^{13}$ discrete
signals per second.    At a finer physical level, it contains 
about $10^{26}$ atoms, and tracking their coordinates (in a 
classical model) would require $10^{78}$ independent parameters.    
As far as we can measure it, the
bit rate of information needed to describe our conscious 
states of mind is far smaller than these numbers. 
For example, lexical decision tasks may take us  
only of the order of $100$ bits per second of
information processing \cite{del2011macroscopic}, not all of which is conscious. 
The number
of details we can attend to per second in visual images also appears
relatively small in comparison.  For example, recognizing one object drawn from a class of 
$\approx 10^5$ takes us of order $100$ms \cite{biederman1987recognition}. 

Admittedly, we cannot be at all precise on this point without a 
precise description of the range of possible conscious
mind states.   Perhaps our emotional spectra are far richer
than we generally credit.  Introspection cannot be precisely
calibrated.   We cannot definitively refute the hypothesis  
that conscious mind states are described by more than
$10^{13}$ binary parameters or even more than $10^{78}$ 
continuous parameters, nor that every neuron firing
or even every atom moving in the brain subtly modulates our 
conscious state.   But these seem very unlikely: most of these parameters seem to be irrelevant to
describing consciousness, since most microscopic details of the 
brain's structure and operation do not seem to have 
conscious correlates.  
So I will take it as a reasonable working hypothesis that 
very many different complete brain states appear to map to the same conscious mind
state, and that we can reasonably model possible human conscious mind
states, as they evolve in time, 
by significantly fewer than $10^{13}$ discrete signals per
second.\footnote{Precise measures of information would need
a probability distribution on the set of possible states.  
In principle, we could get some probability distribution 
on human brain states from 
neuroscientific models derived from observational data 
in typical environments.   We could also get an upper
bound on the information contained in conscious mind
states, if the set of such states is finite,  by assuming they
are equiprobable.}  If so, any physical
theory of consciousness must involve, among other things,
a great deal of data selection.   

Perhaps we can model this as a {\it data
selection principle}: some rule that maps the large amount of 
information contained in a parametrised description of the complete
physical state of the brain to a smaller amount
of information contained in a hypothetical parametrised
description\footnote{
We give a cartoon of such a description below just to 
illustrate one way to begin thinking about possibilities.} 
of the contents of the associated conscious mind.  

I am not sure that anyone currently has any compellingly plausible idea as to how this
might work in any detail.\footnote{Some noteworthy discussions of the
relationship between consciousness and physics include
Refs. \cite{tononi1998consciousness,crick2003framework,tegmark2014consciousness,
oizumi2014phenomenology}.}
Certainly I don't.  
So let me instead give a cartoon: 
not an idea to be taken seriously, but an illustration of 
the sort of thing that would count as a data selection principle.
Suppose that nature has fixed a cubic lattice with a certain
scale $L$, where $L$ is larger than a small molecule and maybe
not much larger than a neuron.\footnote{Any readers who
want to pay this crude cartoon the honour of worrying about its  
consistency with special relativity could imagine that this
story works only where large enough local collections of matter define
a local inertial frame to a good enough approximation.   We could then
replace the assumption of a global lattice with a patchwork of local
lattices, locally defined in the relevant inertial frames.   To extend
the cartoon a bit further we could (for example) add the
rule that if there is no such frame, then there are no qualia, 
and no local consciousness.}
Suppose that nature is described by some version of quantum theory in
which collapses are objectively defined localized events in space-time -- for 
example, some versions of Copenhagen quantum theory, or a dynamical collapse
model.  
Suppose moreover that this version of quantum theory allows us 
to define a wave function on any given spacelike hypersurface,
from some theory of the initial conditions.    
Suppose also that it allows us 
to define local density matrices in a spatial region by 
tracing out from the wave function the degrees of freedom corresponding to other
regions of a spacelike hypersurface in the limiting case where that
hypersurface tends to the past light cone of a region.\footnote{
For example, in Copenhagen quantum theory, this construction means
that the effects of measurements inside the past light cone are taken
into account in constructing the local density matrix, while those
outside the past light cone are not.}   
We then take the expectation value of the mass density
defined by the local density matrix describing the quantum state of 
matter within each volume $L^3$ cube as one component of a 
primitive physical ontology, which describes physical states
to which conscious mind states may be attached. 
We update these density
matrices at each time interval $L/c$, supposing that nature
has also fixed a one-dimensional lattice in time. 

We suppose further that conscious mind states are composed of
combinations of elementary qualia, which supervene on the physical
ontology in a lawlike way.  
Specifically, we suppose there is some local rule according to which a 
quale $Q_j$ is associated with the cube $C$ at discrete 
time point $T$ provided that the configuration of local
density matrices for nearby and recent cubes (within distance $N L$
and within past time $NL/c$, for some number $N > 1$) satisfies
some property $P_j ( \rho_1 , \ldots , \rho_{M} )$.   Here 
$M \approx 8 N^4$ is the number of nearby and recent cubes,
$\rho_i$ are the mass density expectation values in these cubes, 
$j \in J$ is an index over the possible types of quale (which
we might perhaps take to be finite), and the properties $P_j$ are 
sets of mathematical constraints (which to simplify the cartoon we might take to be
exclusive, so that each cube is associated with at most one quale).   
If none of the constraints $P_j$ hold, then
there is no quale associated with the given cube at the given time.  
The consciousness $C(S)$ associated at any given time with a system
$S$ to which these rules are applied is the collection of all
the qualia defined at that time.   

Within our cartoon, this rule is meant
to be fundamental, not tailored to the specifics of human 
brains.   It is supposed to give us a general algorithm 
for identifying $C(S)$ for any system $S$. 
So we really should extend the cartoon to give some cartoon-level story about
how we can tell whether qualia are part of the same consciousness
or not.   Perhaps we could do that by adding a second scale $K$, 
and saying that any pair of qualia separated by no 
more than $K$ cubes, at any given time, form part of the same
consciousness, and that belonging to the same consciousness
is a transitive relation on qualia.  
In other words, a pair of qualia belong to different consciousnesses
if and only if they are not joined by a 
path through the qualia that takes no more than $K$ cubes
for each step.\footnote{The so-called combination problem \cite{seager1995consciousness},
first raised by William James \cite{james2013principles}, is widely thought to be 
an obstacle to any satisfactory theory of consciousness
involving novel rules attaching elements of consciousness
to simple physical system.   Chalmers \cite{chalmers2013combination}
has argued that
the combination problem has several aspects. 
If one thinks of this cartoon as
a sort of panprotopsychist model, the rule that transitively connects
qualia separated by no more than $K$ cubes illustrates 
that panprotopsychist laws (at least at cartoon level) can deal with  
what Chalmers calls the subject combination problem.   
Other aspects, especially the quality combination problem,
remain problematic.}  

For anything like this to work, even at the cartoon level,  
one would have to find properties $P_j$ that tend to be 
correlated with specific conscious qualia when those
properties apply to human brains and (perhaps) central nervous
systems, so that postulating that the qualia supervene
on brain matter gives a good description of our conscious
mind states. The $P_j$ should also not have the property that
these supervenience postulates also imply additional qualia from
(at the very least) most of the matter surrounding our brains
and central nervous systems: if they did, we would not be able to speak of separate 
single consciousnesses associated with each brain.   One would also need some
plausible description of elementary qualia $j \in J$.
And then, much harder still, one would need that the $P_j$
actually produce the right sort of collections of qualia 
-- corresponding to the sort of things we actually consciously
experience -- for the enormous variety of brains and brain states
for which we have experience (direct or reported). 
It does not matter for the cartoon whether or not the $P_j$  
imply that things other than brains -- modern computers, 
large rocks, spiral nebulae -- are also conscious.  

Obviously, I am not suggesting any of this is actually possible. 
The aim of this sketch is not to speculate about an actual theory of
consciousness but just to give a concrete, albeit incredible,
illustration that allows us to develop a particular line of thought further. 
Whatever the fundamental physical theory of consciousness -- 
if there is one -- looks like, I am pretty sure it does not
resemble this cartoon.   
But suppose, just for the sake of the argument, that it 
were possible to make the cartoon work.
We would then have a theory of consciousness, including a  
classification of qualia and a data selection principle.
The description of the $P_j$ would, we need to assume, be 
significantly simpler than just a dictionary of all the 
brain states and corresponding conscious states that we
can identify.   (If it isn't, then it doesn't produce a compressed
description of the empirical data about consciousness, and so 
it doesn't define a useful theory.)  In that sense, we would have a significantly
better understanding of consciousness.\footnote{This is not to say
we would have solved the existing puzzles concerning the relationship
of consciousness and physics.   Indeed, a theory of this type 
would raise new puzzles: for example, why the lawlike
supervenience of qualia on a quantum-derived ontology takes
this particular form.   Nonetheless, we would be able to 
make predictions we cannot presently make about the 
conscious mind states associated with brain states.
We could also test them, insofar as any predictions about
conscious states are testable, by adding 
the assumption that conscious states are often reliably reported. 
That is, we could see whether
the predicted conscious mind states agree with our descriptions of our own mind states 
and with the descriptions others give of theirs.}

However, our theory, as described, would be of consciousness
as an epiphenomenon.   It could possibly nonetheless
represent a very substantial advance in our 
understanding of consciousness, if it turned out to 
describe the rich variety of our experiences from
a simple set of principles $P_j$.    But it could not 
explain how and why humans had evolved to produce brains
that just happen to produce conditions in which many 
$P_j$ tend to apply, and in which the corresponding qualia
produce the sort of consciousnesses we have. 
So, on the anti-epiphenomenal view we have outlined above,
it could not be fundamentally correct: at best it might
be a good approximation.    

\subsection{Improving the cartoon?}

Any explanation of why humans and other animals evolved to
become conscious has to run one of two ways.   

One is that human evolution can be understood purely in terms
of the familiar material laws of physics, and it is just a nice property
of consciousness that it resides in highly evolved creatures
that are continually processing information about their
environments and acting on it.   If one believes this is 
a satisfactory definition of, or a self-evident property of,
consciousness, one can be happy with this explanation.
As noted above, I don't, so I'm not.  

The other is that familiar materialist explanations of evolution alone
are not adequate and that something about consciousness itself gives an
extra evolutionary advantage.   This needs an extra mechanism 
that implies that, in some sense, conscious creatures tend to
prevail in competition with unconscious ones.   More than that,
since a binary division between conscious and unconscious creatures 
doesn't give enough room for an evolutionary story, it needs to imply,
in some sense, that more conscious creatures tend to prevail in
competion with less conscious ones.   

Having (perhaps foolhardily!) chosen to reject the first type of
explanation in this discussion, we have to try for the second.   
We can translate 
``more (less) conscious'' into ``having more (fewer) qualia'' in
our cartoon.   
Then, fortunately for our cartoon narrative, there 
is at least an available option, already explored in a 
different connection \cite{kent2013beable} as a natural way of 
defining generalizations of quantum theory.   
According to our cartoon, we can (in principle) calculate the 
probability $P_q (D)$ of any distribution $D$ of qualia, from 
quantum dynamics and the relevant measurement or collapse postulate 
and from knowledge of the constraints defining
the properties $P_j$.   (The suffix $q$ stands for quantum here.)  We can do this for any
system $S$, or in principle (given a good enough quantum theory 
that incorporates gravity and describes cosmology)  
for the entire universe.   
As noted, if our cartoon were actually
correct, this calculation would give the correct predictions
for an epiphenomenal
model of consciousness.     
But we can change the model, and make it {\it non-epiphenomenally}
dependent on quantum theory, if we postulate instead that the 
true probability distribution $P_{\rm true} (D) $ of distributions of 
qualia is a modified version of $P_q (D)$. 

For instance, following the ideas of Ref. \cite{kent2013beable},  
we could postulate that 
\begin{equation}\label{qualiapostulate}
P_{\rm true} (D) = C P_q (D) A(D) \, ,
\end{equation} 
where $C$ is a constant that ensures the rescaled probabilities sum to
$1$ and $A(D)$ is some weight factor that depends only on properties
of the qualia distribution $D$.\footnote{This is not the most general
possibility, but general enough for the present discussion.}
To be clear: if we take quantum theory as ultimately a theory for predicting 
the experiences of observers, this means postulating that quantum
theory is at least subtly incorrect.    But the deviation could be 
very small and subtle, if $A(D)$ depends only slightly and subtly on $D$.  

Now, if $A(D)$ is chosen to favour, even very slightly, distributions with
more qualia, we have the potential beginnings of an explanation for the evolution of 
primate-level consciousness from primitive qualia.   
For such an explanation to work, we need that the
postulated properties $P_j$
somehow just happen to involve relations among mass density
expectation values that are useful
for, or naturally fit into the context of, the types of information
processing that animal brains carry out -- gathering information 
correlated with their bodies and environments, computing relevant
features, and generating responses.\footnote{``Information'' here is 
meant in the slightly informal sense standardly used
in discussing biological intelligence.   It is very hard to quantify
precisely the information in an animal's environment, or its representation
of that environment, or its behavioural responses.   Nonetheless it is generally
agreed that relatively simple information-theoretic models give us
good analogies, and widely expected that the analogies could in principle be made to   
approach representations of reality more and more closely as more
complexity is introduced.   Underlying this expectation is the
assumption -- with which most neuroscientists and physicists are very
comfortable -- that the known laws of physics completely describe animal behaviour.}  

But given that (big \footnote{Though it {\it is} a very big 
  assumption to make, it is still much less contrived than the
  dogmatic assumption that we 
  must have consciousnesses of {\it exactly} the sort we have, given the
  information processing that we do.}) assumption, we can see that there would be selection
pressure towards creatures whose information processing capacities
use such relations in their information processing systems, and 
then selection pressure in favour of those whose systems
generate more qualia.\footnote{
Stretching credulity even further, 
if $A(D)$ were somehow chosen to favour  qualia of particular types
(which tend to be ``pleasant'') and disfavour qualia of other types
(which tend to be ``painful''), we might also have at least the potential
beginning of a story about how creatures came to embed in their 
information processing systems some subsystems that generate 
pleasant qualia (which are favoured by our hypothetical
postulate, and which are located so that they correspond to 
evolutionarily favourable activities) and some that generate
painful qualia (which are disfavoured, and located so that they
correspond to unfavourable activities).   
But our comments earlier
apply: there seems no reason why this would not equally well work
for evolution  -- although not so happily for us, its conscious products
-- with the pleasure-pain polarities reversed.   
The best I can offer is the thought that that the pleasure-pain
problem might somehow look different
and less fundamentally threatening if we understood the actual details of 
the interaction between material states and conscious states.
But we don't have a theory of consciousness, and so I 
don't see how this could work.   
Maybe, of course, it just doesn't: maybe 
the pleasure-pain problem actually is insoluble in this approach, or
perhaps in any approach.
} 
 
One can only make full sense of this cartoon theory as we have 
phrased it, with a postulate of the
form (\ref{qualiapostulate}), 
in a block universe picture, since equation (\ref{qualiapostulate}) 
defines the probability distribution for the complete configuration
of all qualia throughout space and time.\footnote{This leaves 
us with the problem of the psychological perception of time,
since those qualia must  
nonetheless give conscious creatures the impression of a flow
of time associated with a succession of experiences. 
But this is a deep problem in any view of consciousness and physics.
It also, of course, leaves us with the usual gap between 
standard language and fundamental ontology  
in talking about processes within a block universe. 
For example, ``the evolution of primate-level consciousness'' 
is shorthand for something like ``the sequential appearance of
low-level, increasingly more complex, and primate-level consciousnesses
at increasing cosmological times measured from the presumably
highly ordered singularity conventionally referred to as the beginning
of the universe''.}
Block universe theories of this type are logically consistent, but
they can have unusual and counter-intuitive implications, including
effects that appear to agents within the theory to be reverse 
causation and spacelike signalling. 
They also do not generally reduce to equally simple theories 
applicable to subsystems of the universe.
For example, the behaviour of a conscious individual, or an 
finite ecosystem, cannot generally be modelled using only
its initial state and some simple analogue of 
Eqn. (\ref{qualiapostulate}). 
There {\it are} nonetheless theoretical reasons to consider some types
of block universe theory, since they suggest possible solutions to the
quantum measurement/reality problem. \footnote{One line of 
thought on this can be found in Refs. \cite{kent2015lorentzian,kent2016quantum};
see also references therein for other discussions.} 
That said, like
the earlier part of the cartoon, our block universe qualia cartoon theory is meant only as an 
existence theorem, not a serious theoretical 
proposal.\footnote{Among the very odd features of the block universe rule
  (\ref{qualiapostulate}), as stated, is that it implies that the bias 
towards consciousness in evolutionary selection arises from a
calculation
that depends on the global distribution of consciousness in space and
time.  For example, the weight bias between alternatives that would produce either 
descendent $d_1$ or $d_2$, with differing numbers of lifetime qualia
does not in general reduce to a ratio of the weights associated with
the qualia they produce over their lifetimes,  $A(C( d_1) )/A( C(
d_2))$.  It does not even reduce to a ratio of the form  $A(T_1 )/
A(T_2)$, defined by the weights associated 
with the qualia their entire trees of descendents $T_i$ produce.  One has also to consider the effects of $T_1$ and $T_2$
on other future conscious lifeforms and evaluate the ratio of weights
$A(D_1)/A(D_2)$ associated with the full qualia distributions
$D_i$ over all future space-time arising if $d_i$ is the descendent. 
A version of the rule in which the weight function
$A$ is chosen so that $A(D_1 ) / A(D_2) = A(C (d_1 ))/  A(C(d_2))$ 
would probably improve the cartoon somewhat.}

\section{Summary and discussion}

Every line of thought on the relationship of consciousness to physics 
runs into deep trouble.  Because of this, we are inclined to place 
some (albeit weak) credence that the line of thought we have outlined
may not be entirely orthogonal to the truth, despite its own evident problems.
We stress again that none of the details of our cartoons are meant to be taken seriously.
What we do take seriously, at a weak level of credence, is  
the suggestion that we could make some progress on understanding
the problem of the evolution of consciousness if we supposed that 
consciousnesses alter (albeit perhaps very slightly and subtly)
quantum probabilities.   A further reason for taking this seriously
(still at a weak level of credence) is an aesthetic preference
for theories in which fundamental quantities (here qualia and quanta)
genuinely interact, rather than one being purely dependent on the
other.   The same point was used to motivate inventing and testing 
generalizations of quantum theory in a different context in
Ref. \cite{kent2013beable}.\footnote{There are interesting parallels between the
beable hypothesis \cite{bell1976theory} and the qualia hypothesis.    
Bell's notion of beable -- a mathematical quantity in a physical
theory that directly corresponds to an element of physical reality -- 
is intended to give a language to highlight a general class of possible
solutions to the quantum reality problem: what, precisely, could
be the sample space for which we calculate probabilities for a 
closed quantum system?    Similarly, the notion of qualia identifies
hypothetical elementary quantities that could be characterised by a theory
of consciousness that addresses the hard problem.        
In both cases, whether the relevant problem really {\it is} a problem 
is controversial, with thoughtful people on both sides. 
For those who think both problems are real and may lead to new
science, parsimony might suggest that qualia can be understood in
terms of -- perhaps as functions of -- beables.   A more radical 
option would be to identify beables and qualia, which would imply
that quantum theory is ultimately about conscious perceptions 
(though not necessarily only those of familiar living creatures). }

What are the implications?   Broadly, to add some support to tests of 
quantum theory that involve conscious observers.   
For example, perhaps this line of thought adds a little to the motivation for interferometry experiments involving
viruses \cite{arndt2002interferometry}, or ultimately bacteria or
larger creatures.\footnote{Among the further uncertainties here are that 
it is far from clear why we should expect viruses or bacteria to have 
any sort of consciousness.}   
Existing intuitions that such experiments might be worthwhile
are mostly based on the idea \cite{wigner1961remarks} that quantum collapse may be 
connected to, or even directly caused by, consciousness.   Chalmers and McQueen \cite{chalmersfqxi}
have recently formulated a more precise version of this proposal,
invoking the hypothesis of Tononi and collaborators
\cite{oizumi2014phenomenology}
that what they term integrated information
could define a measure of consciousness.\footnote{To the best of my
knowledge there is currently no empirical evidence for the integrated
information hypothesis, and its plausibility is debated.   
However, Chalmers and McQueen's point applies quite
generally.   Any well-defined measure of consciousness would allow a
more precise formulation of Wigner's idea and hence an analysis of
whether experimental tests could be feasible in the forseeable future.}
Our discussion gives another weak reason  
for speculating that the direct involvement of conscious observers 
might possibly alter something relevant to interferometry and other
experiments.

Our discussion perhaps also adds a little to the motivation for long range Bell experiments in 
which human observers make (their best attempt at) free random
choices of measurement outcomes, and observe the outcomes directly, with separations large enough that 
the combined choice processes and observations on the two wings are
spacelike separated.   This added motivation is presently weak,
since we have given no specific motivation in this discussion for looking at Bell
experiments in particular.
A perhaps stronger motivation comes from
combining the hypotheses that wave function collapse requires
consciousness and that collapse results propagate causally in
the future light cone.   This leads to a consistent theory, if
one assumes either that measurements can never be precisely
specified or that collapses are never perfect projections, and
implies a loophole (the so-called ``collapse locality loophole'') 
in all Bell experiments to date \cite{kent2005causal,salart2008spacelike}. 

(It is worth parenthetically mentioning here that some \cite{hardy2017,big2018challenging} have suggested that Bell
experiments involving conscious observers can also be motivated 
by some form of ``free will'' hypothesis.   
Discussions of free will in connection with physics are, if anything, even more
contentious than those of consciousness, and it is beyond my scope to
try to add to them here, beyond pointing readers to recent
relevant work.   
Roughly speaking, as I understand it, the main ideas are that
(a) superdeterminism could explain the observed violation of Bell
inequalities in Bell experiments to date, (b) there are possible
motivations for considering models in which superdeterminism 
applies to the material world but not to the outcomes of freely made human decisions. 
One such motivation, discussed by Hardy \cite{hardy2017} is some form of dualism, in which
human decisions have the effect of unpredictable interventions into
the material world, whose effects propagate into but not outside 
the future light cone of the decision point.   
On this view, it is possible we might
see different results in Bell experiments in which
the measurements on the two wings arise from spacelike separated free
choices by observers.   Retarded Bell inequalities formalising 
mathematically the hypothesis to be tested were defined by
Hardy \cite{hardy2015bell}.  An extended discussion is given in 
Ref. \cite{hardy2017}, where Hardy reviews the history of work and ideas
in this direction and sets out a detailed experimental
proposal, while also noting problems with and arguments against the relevant hypotheses. 
While Hardy expresses strong credence that Bell experiments will continue to
give standard results, he stresses the point, also made below, that
the payoff of a surprising result is sufficiently large to justify 
the experiments.  

Arguments that free will should play a fundamental role 
in physics have also recently been made by Gisin
\cite{gisin2016time}.) 

In the longer term, if and when quantum technology advances to the
point that direct tests of quantum theory (not necessarily interferometric tests) on macroscopic objects are 
possible, our discussion does give a clear motivation for carrying them out on
animals and humans. 

To be clear, neither I nor (as far as I am aware) any author
mentioned here insist or even predict that quantum theory will be violated in any of these
experiments.  The arguments that consciousness might have a role in 
quantum physics are admittedly problematic, even if the
counterarguments also are.   And most of the interesting theoretical
ideas about quantum theory over the last fifty years involve
formulations in which observers play no special role.  

But, to be provocatively quantitative, on the grounds that deep puzzles
in physics have often led to big surprises and that consensus views tend to 
be overconfident, I would still give credence of perhaps $15 \% $ that 
{\it something} specifically to do with consciousness
causes deviations from quantum theory, with perhaps $3 \%$ credence that 
this will be experimentally detectable within the next fifty years.  
No doubt many physicists would give much lower figures.  
Still -- as the existential risk community in particular 
has emphasized (e.g. \cite{,kent2004critical,bostrom2011global,bostrom2013existential}) -- if one assigns non-zero
probabilities, however small and uncertain, to events with large costs or benefits,
one should pay close attention to the expectation values.   
The potential benefits here include making 
some progress in understanding the relationship between physics and consciousness.  
That would also offer some hope of getting data to guide
us in the ethical questions we already face (how rich are the  
consciousnesses of animals?) and those we likely will (are human-level 
AI programmes, or human brain emulations?).   
It could also significantly change our understanding of the physics
of computation, with potentially large implications for
the future of intelligence.   
Even if one has very weak levels of credence (say $0.01 \%$) for any current ideas on the physics of 
consciousness, it seems to me the large potential implications still suggest 
that more work should be carried out on possible experiments on conscious or plausibly
conscious observers, and their possible theoretical motivations.

\section{Acknowledgements}
This work was supported by an FQXi grant and by Perimeter Institute
for Theoretical Physics. Research at Perimeter Institute is supported
by the Government of Canada through Industry Canada and by the
Province of Ontario through the Ministry of Research and Innovation.
I thank David Chalmers, Lucien Hardy and Graeme Mitchison for helpful and enjoyable discussions.  

\section*{References}

\bibliographystyle{unsrtnat}
\bibliography{qualia2}{}

\begin{thebibliography}{45}
\providecommand{\natexlab}[1]{#1}
\providecommand{\url}[1]{\texttt{#1}}
\expandafter\ifx\csname urlstyle\endcsname\relax
  \providecommand{\doi}[1]{doi: #1}\else
  \providecommand{\doi}{doi: \begingroup \urlstyle{rm}\Url}\fi

\bibitem[James(1879)]{jamesautomata}
William James.
\newblock Are we automata?
\newblock \emph{Mind}, 4:\penalty0 1--22, 1879.

\bibitem[Chalmers(2006)]{Chalmers2006-CHASAW}
David~J. Chalmers.
\newblock Strong and weak emergence.
\newblock In P.~Davies and P.~Clayton, editors, \emph{The Re-Emergence of
  Emergence}. Oxford University Press, 2006.

\bibitem[Gennaro(2018)]{iepconscious}
Rocco~J. Gennaro.
\newblock Consciousness.
\newblock \emph{The Internet Encyclopedia of Philosophy}, 2018.
\newblock URL \url{http://www.iep.utm.edu/}.

\bibitem[Chalmers(1996)]{chalmers1996conscious}
David~J Chalmers.
\newblock The conscious mind: In search of a fundamental theory, 1996.

\bibitem[Nida-R\"umelin(2015)]{knowledge}
Martine Nida-R\"umelin.
\newblock Qualia: The knowledge argument.
\newblock \emph{The Stanford Encyclopedia of Philosophy (Summer 2015 Edition)},
  2015.
\newblock URL
  \url{https://plato.stanford.edu/archives/sum2015/entries/qualia-knowledge/}.

\bibitem[Alter(2017)]{iepknowledge}
Torin Alter.
\newblock The knowledge argument against physicalism.
\newblock \emph{The Internet Encyclopedia of Philosophy}, 2017.
\newblock URL \url{http://www.iep.utm.edu/}.

\bibitem[Robinson(2015{\natexlab{a}})]{stanepi}
William Robinson.
\newblock Epiphenomenalism.
\newblock \emph{The Stanford Encyclopedia of Philosophy (Fall 2015 Edition)},
  2015{\natexlab{a}}.
\newblock URL
  \url{https://plato.stanford.edu/archives/fall2015/entries/epiphenomenalism/}.

\bibitem[Chalmers(1995)]{chalmers1995facing}
David~J Chalmers.
\newblock Facing up to the problem of consciousness.
\newblock \emph{Journal of consciousness studies}, 2\penalty0 (3):\penalty0
  200--219, 1995.

\bibitem[Dennett(1993)]{dennett1993consciousness}
Daniel~C Dennett.
\newblock Consciousness explained, 1993.

\bibitem[Glynn(2003)]{glynn2003anatomy}
Ian Glynn.
\newblock \emph{An anatomy of thought: The origin and machinery of the mind}.
\newblock Oxford University Press, 2003.

\bibitem[Weisberg(2017)]{iephard}
Josh Weisberg.
\newblock The hard problem of consciousness.
\newblock \emph{The Internet Encyclopedia of Philosophy}, 2017.
\newblock URL \url{http://www.iep.utm.edu/hard-con/}.

\bibitem[Walter(2017)]{iepepi}
Sven Walter.
\newblock Epiphenomenalism.
\newblock \emph{The Internet Encyclopedia of Philosophy}, 2017.
\newblock URL \url{http://www.iep.utm.edu/epipheno/}.

\bibitem[Van~Gulick(2018)]{sep-consciousness}
Robert Van~Gulick.
\newblock Consciousness.
\newblock In Edward~N. Zalta, editor, \emph{The Stanford Encyclopedia of
  Philosophy}. Metaphysics Research Lab, Stanford University, spring 2018
  edition, 2018.

\bibitem[Robinson(2015{\natexlab{b}})]{stanepiselfstult}
William Robinson.
\newblock Epiphenomenalism: Self-stultification.
\newblock \emph{The Stanford Encyclopedia of Philosophy (Fall 2015 Edition)},
  2015{\natexlab{b}}.
\newblock URL
  \url{https://plato.stanford.edu/entries/epiphenomenalism/#SelStu}.

\bibitem[Lewis(1929)]{lewis1929mind}
Clarence~Irving Lewis.
\newblock \emph{Mind and the world-order: Outline of a theory of knowledge}.
\newblock Courier Corporation, 1929.

\bibitem[Schr{\"o}dinger(1992)]{schrodinger1992life}
Erwin Schr{\"o}dinger.
\newblock \emph{What is life?: With mind and matter and autobiographical
  sketches}.
\newblock Cambridge University Press, 1992.

\bibitem[del Prado~Mart{\'\i}n(2011)]{del2011macroscopic}
Ferm{\'\i}n~Moscoso del Prado~Mart{\'\i}n.
\newblock Macroscopic thermodynamics of reaction times.
\newblock \emph{Journal of Mathematical Psychology}, 55\penalty0 (4):\penalty0
  302--319, 2011.

\bibitem[Biederman(1987)]{biederman1987recognition}
Irving Biederman.
\newblock Recognition-by-components: a theory of human image understanding.
\newblock \emph{Psychological review}, 94\penalty0 (2):\penalty0 115, 1987.

\bibitem[Kent(2013)]{kent2013beable}
Adrian Kent.
\newblock Beable-guided quantum theories: Generalizing quantum probability
  laws.
\newblock \emph{Physical Review A}, 87\penalty0 (2):\penalty0 022105, 2013.

\bibitem[Arndt et~al.(2002)Arndt, Nairz, and
  Zeilinger]{arndt2002interferometry}
Markus Arndt, Olaf Nairz, and Anton Zeilinger.
\newblock Interferometry with macromolecules: Quantum paradigms tested in the
  mesoscopic world.
\newblock In \emph{Quantum [Un] Speakables}, pages 333--350. Springer, 2002.

\bibitem[Wigner(1961)]{wigner1961remarks}
Eugene~P Wigner.
\newblock Remarks on the mind-body problem.
\newblock In I.~J. Good, editor, \emph{The Scientist Speculates}. Heineman,
  1961.

\bibitem[Chalmers(2016)]{chalmersfqxi}
D.~Chalmers.
\newblock Dirty secrets of consciousness.
\newblock \emph{Talk at FQXi 5th International Conference, Banff, August 2016},
  2016.

\bibitem[Oizumi et~al.(2014)Oizumi, Albantakis, and
  Tononi]{oizumi2014phenomenology}
Masafumi Oizumi, Larissa Albantakis, and Giulio Tononi.
\newblock From the phenomenology to the mechanisms of consciousness: integrated
  information theory 3.0.
\newblock \emph{PLoS Comput Biol}, 10\penalty0 (5):\penalty0 e1003588, 2014.

\bibitem[Kent(2005)]{kent2005causal}
Adrian Kent.
\newblock Causal quantum theory and the collapse locality loophole.
\newblock \emph{Physical Review A}, 72\penalty0 (1):\penalty0 012107, 2005.

\bibitem[Salart et~al.(2008)Salart, Baas, van Houwelingen, Gisin, and
  Zbinden]{salart2008spacelike}
Daniel Salart, Augustin Baas, Jeroen~AW van Houwelingen, Nicolas Gisin, and
  Hugo Zbinden.
\newblock Spacelike separation in a {B}ell test assuming gravitationally
  induced collapses.
\newblock \emph{Physical review letters}, 100\penalty0 (22):\penalty0 220404,
  2008.

\bibitem[Hardy(2017)]{hardy2017}
L.~Hardy.
\newblock Proposal to use humans to switch settings in a {B}ell experiment.
\newblock \emph{arXiv preprint arXiv:1705.04620}, 2017.

\bibitem[{B}ell Test~Collaboration et~al.(2018)]{big2018challenging}
BIG {B}ell Test~Collaboration et~al.
\newblock Challenging local realism with human choices.
\newblock \emph{arXiv preprint arXiv:1805.04431}, 2018.

\bibitem[Hardy(2015)]{hardy2015bell}
L.~Hardy.
\newblock {B}ell inequalities with retarded settings.
\newblock \emph{arXiv preprint arXiv:1508.06900}, 2015.

\bibitem[Gisin(2015)]{gisin2016time}
N.~Gisin.
\newblock Time really passes, science can't deny that.
\newblock \emph{arXiv preprint arXiv:1602.01497}, 2015.

\bibitem[Kent(2004)]{kent2004critical}
Adrian Kent.
\newblock A critical look at risk assessments for global catastrophes.
\newblock \emph{Risk Analysis}, 24\penalty0 (1):\penalty0 157--168, 2004.

\bibitem[Bostrom and Cirkovic(2011)]{bostrom2011global}
Nick Bostrom and Milan~M Cirkovic.
\newblock \emph{Global catastrophic risks}.
\newblock Oxford University Press, 2011.

\bibitem[Bostrom(2013)]{bostrom2013existential}
Nick Bostrom.
\newblock Existential risk prevention as global priority.
\newblock \emph{Global Policy}, 4\penalty0 (1):\penalty0 15--31, 2013.

\bibitem[Saunders et~al.(2010)Saunders, Barrett, Kent, and
  Wallace]{saunders2010many}
Simon Saunders, Jonathan Barrett, Adrian Kent, and David Wallace.
\newblock \emph{Many Worlds?: Everett, Quantum Theory, \& Reality}.
\newblock OUP Oxford, 2010.

\bibitem[Schlosshauer(2007)]{schlosshauer2007decoherence}
Maximilian~A Schlosshauer.
\newblock \emph{Decoherence: and the quantum-to-classical transition}.
\newblock Springer Science \& Business Media, 2007.

\bibitem[Fuchs(2003)]{fuchs2003notes}
Christopher~A Fuchs.
\newblock Notes on a {P}aulian idea: Foundational, historical, anecdotal, and
  forward-looking thoughts on the quantum: Selected correspondence, 1995-2001,
  2003.

\bibitem[Cushing et~al.(2013)Cushing, Fine, and Goldstein]{cushing2013bohmian}
James~T Cushing, Arthur Fine, and Sheldon Goldstein.
\newblock \emph{Bohmian mechanics and quantum theory: an appraisal}, volume
  184.
\newblock Springer Science \& Business Media, 2013.

\bibitem[Tononi and Edelman(1998)]{tononi1998consciousness}
Giulio Tononi and Gerald~M Edelman.
\newblock Consciousness and complexity.
\newblock \emph{science}, 282\penalty0 (5395):\penalty0 1846--1851, 1998.

\bibitem[Crick and Koch(2003)]{crick2003framework}
Francis Crick and Christof Koch.
\newblock A framework for consciousness.
\newblock \emph{Nature neuroscience}, 6\penalty0 (2):\penalty0 119--126, 2003.

\bibitem[Tegmark(2014)]{tegmark2014consciousness}
Max Tegmark.
\newblock Consciousness as a state of matter.
\newblock \emph{arXiv preprint arXiv:1401.1219}, 2014.

\bibitem[Seager(1995)]{seager1995consciousness}
William Seager.
\newblock Consciousness, information and panpsychism.
\newblock \emph{Journal of Consciousness Studies}, 2\penalty0 (3):\penalty0
  272--288, 1995.

\bibitem[James(2013)]{james2013principles}
William James.
\newblock \emph{The principles of psychology}.
\newblock Read Books Ltd, 2013.

\bibitem[Chalmers(2013)]{chalmers2013combination}
David~J Chalmers.
\newblock The combination problem for panpsychism.
\newblock \emph{Panpsychism: Contemporary Perspectives}, 2013.

\bibitem[Kent(2015)]{kent2015lorentzian}
Adrian Kent.
\newblock Lorentzian quantum reality: postulates and toy models.
\newblock \emph{Phil. Trans. R. Soc. A}, 373\penalty0 (2047):\penalty0
  20140241, 2015.

\bibitem[Kent(2016)]{kent2016quantum}
Adrian Kent.
\newblock Quantum reality via late time photodetection.
\newblock \emph{arXiv preprint arXiv:1608.04805}, 2016.

\bibitem[Bell(1976)]{bell1976theory}
John~S Bell.
\newblock The theory of local beables.
\newblock \emph{Epistemological Letters}, 9\penalty0 (11), 1976.

\end{thebibliography}
\end{document}